\documentclass[journal,10pt]{IEEEtran} 
\ifCLASSINFOpdf
\usepackage{graphicx}
\DeclareGraphicsExtensions{.pdf}
\else
\usepackage[dvips]{graphicx}
\fi
\DeclareGraphicsExtensions{.eps}
\usepackage{amsmath}
\usepackage{cite}
\usepackage{xcolor}
\usepackage{siunitx}
\DeclareSIUnit\bps{bps}
\DeclareSIUnit\Torr{Torr}
\usepackage{wrapfig}
\usepackage{epstopdf}
\usepackage{cases}
\usepackage[tight,footnotesize]{subfigure}
\usepackage{amssymb}
\usepackage{verbatim}
\normalsize

\begin{document}


\title{ Information Rates of Controlled Protein Interactions Using Terahertz  Communication}
 \author{Hadeel Elayan, Andrew W. Eckford, and Raviraj Adve 


\thanks{We would like to acknowledge the support of the National Science and Engineering Research Council, Canada, through its Discovery Grant program.}
\thanks{H. Elayan and R. Adve are with the Edward S. Rogers Department of Electrical and Computer Engineering, University of Toronto, Ontario, Canada, M5S 3G4 (e-mail: hadeel.mohammad@mail.utoronto.ca; rsadve@ece.utoronto.ca).}

\thanks{ A. Eckford is  with the Department of Electrical Engineering and Computer Science, York University, Ontario, Canada, M3J 1P3 (e-mail: aeckford@yorku.ca).}}




\maketitle
\begin{abstract} 
In this work, we present a paradigm bridging electromagnetic (EM) and molecular communication through  a stimuli-responsive intra-body model. It has been established that protein molecules, which play a key role in governing cell behavior, can be selectively stimulated using Terahertz (THz) band frequencies. By triggering protein vibrational modes using THz waves, we induce changes in protein conformation, resulting in the activation of a controlled cascade of biochemical and biomechanical events. To analyze such an interaction, we formulate a communication system composed of a nanoantenna transmitter and a protein receiver. We adopt a Markov chain model to account for protein stochasticity with transition rates governed by the nanoantenna force. Both two-state and multi-state protein models are presented to depict different biological configurations. Closed form expressions for the mutual information of each scenario is derived and  maximized to find the capacity between the input nanoantenna force and the protein state. The results we obtain indicate that controlled protein signaling provides a communication platform for information transmission between the nanoantenna and the protein with a clear physical significance. The analysis reported in this work should further research into the EM-based control of protein networks.    
%
%

\end{abstract}

\section{Introduction}
Interest in nanoscale robotic systems  has led researchers to investigate different frameworks to initiate reliable communication between nanomachines. One solution is molecular communication, which is a  paradigm inspired by nature, that entails utilizing chemical signals as carriers of information. The transmitter of this diffusion-based channel releases  particles into an aqueous or gaseous medium, where the particles propagate until they arrive at the receiver; the receiver then detects and decodes the information in these particles~\cite{farsad2016comprehensive,srinivas2012molecular,pierobon2011diffusion}. As another solution, the emergence of plasmonic nanoantennas has paved the way towards electromagnetic (EM) communication among nanodevices, where both the Terahertz (THz) band~\cite{7955066,7086348,elayan2017photothermal,elayan2018end} and optical frequency range~\cite{johari2018nanoscale} are possible candidates. Specifically, in-vivo wireless nanosensor networks (iWNSNs) have emerged to provide fast and accurate disease diagnosis and treatment. These networks are expected to operate inside the human body in real time while establishing reliable wireless transmission among nanobiosensors~\cite{shubair2015vivo}.


One active research topic within  molecular communications involves establishing interfaces to connect the molecular paradigm with its external environment~\cite{kisseleff2017magnetic, 8467351, liu2017using, krishnan2018wireless}. The authors in~\cite{kisseleff2017magnetic} proposed a wearable magnetic nanoparticle detector to be used as an interface between a molecular communication system deployed inside the human body and a signal processing unit located outside. In~\cite{8467351}, the authors presented a biological signal conversion interface which translates an optical signal into a chemical one by changing the pH of the environment. Moreover,  a redox-based experimental platform has been introduced in~\cite{liu2017using} to span the electrical and molecular domains. This wet-lab coupling paves the way towards novel generation of bio-electronic components that serve as the basis of intelligent drugs, capable of biochemical and electrical computation and actuation. Furthermore, in a very recent work, the authors in~\cite{krishnan2018wireless}, identified genes that control cellular function upon responding to EM fields that penetrate deep tissue non-invasively. Their experimental results complement the growing arsenal of technologies dedicated to the external control of cellular activity in-vivo.

Among the biological structures found in the human body, protein molecules are heterogeneous chains of amino acids; they perform their biological function by coiling and folding into a distinct three dimensional shape as required. Changes in protein level, protein localization, protein activity, and protein-protein interactions are critical aspects of an inter-cellular communication process collectively known as {\em signal transduction}. One important feature associated with protein structures is that their vibrational modes are found in the THz frequency range~\cite{turton2014terahertz}. These modes provide information about protein conformational  change, ligand binding and oxidation state~\cite{knab2006hydration}. Therefore, by triggering protein vibrational modes using THz EM waves, we can direct mechanical signaling inside protein molecules, in turn controlling changes in their structure and, as a result, activating associated biochemical events~\cite{matellan2018no}.

In this work, we bridge the gap between EM (specifically, THz radiation) and molecular communication; We consider a communication link which consists of a nanoantenna transmitter, a protein receiver and a Markovian signal transduction channel. We are interested especially in the process at the receiving end of signal transduction, where a protein changes conformation due to the induced THz signal. Since this problem can be thought of fundamentally as an information transmission problem, our aim in this paper is to compute the mutual information of this communication link. In fact, gaining a detailed understanding of the input-output relationship in biological systems requires quantitative measures that capture the interdependence between components. Hence, a closed form expression for the mutual information rate under independent, identically distributed (IID) inputs is derived and maximized to find the capacity for different protein interaction scenarios. By finding the mutual information rate, experimenters are guided into the amount of information the protein signaling pathway carries. 
 
The main contributions of the paper are as follows:\begin{itemize}
\item 
We model the stochastic protein dynamics actuated through THz waves as a discrete-time, finite-state channel. We present both a two-state and a multi-state model to emulate protein dynamics. In the two-state model, a change in the protein state is triggered through the applied nanoantenna THz force. In the multi-state model, a cascade of changes in the protein configuration is stimulated, where links between different protein states are controlled through the targeted application of THz force.  
\item We analytically derive the mutual information and compute the capacity under different constraints for the two-state and multi-state protein models. The achieved theoretical rates indicate the existence of a ubiquitous mechanism for information transmission between the nanoantenna and the protein with a clear physical significance. 
\end{itemize} 

Biological systems can be generally modelled with microstates; this could refer to the covalently modified state, conformational state, cellular location state, etc. Each of these states defines a certain attribute related to either the protein structure or function~\cite{duan2002describing}. In  our work, the biological meaning of state refers to the conformational state, which we consider as either  Unfolded or Folded for the two-state model. In the case of the multi-state model,  we refer to multiple intermediate states. An example  is the photoactive membrane protein, \textit{Bacteriorhodopsin}. The cycle of this protein consists of several states including a resting state and  a series of photo-intermediate states, each  of which is associated with a  conformational change~\cite{markelz2008terahertz}. The transition between protein states regulates biological processes, including cell signaling. Thereafter, the methodology presented in this work sheds light on various opportunities that impact applications concerning drug discovery, biosensing as well as disease control and prevention.

The rest of the paper is organized as follows. In Sec.~\ref{Sec2}, the system model of the  stimulated protein signal transduction pathway is presented. In Sec.~\ref{Sec3}, a communication system based on Markov finite-states is developed to capture protein dynamics. In Sec.~\ref{Sec4}, a two-state protein model is formulated. The model is further extended and generalized to take into account multi-state protein interactions in Sec.~\ref{Sec5}.  In Sec.~\ref{Sec7}, the numerical results of the models are illustrated while providing a clear physical insight. Finally, we draw our conclusions in Sec.~\ref{Sec8}.

\section{System Model}
\label{Sec2}
\subsection{The Physical Process}
 Living cells communicate with each other through  a series of biochemical interactions referred to as signal transduction networks. A molecular process referred to as mechanotransduction, governs the transmission of mechanical signals from the extracellular matrix to the nucleus~\cite{martino2018cellular}. Proteins, which  are considered major drivers of signal transduction, display a status change in response to mechanical stimulation. In our work, we consider a mechanotransduction communication channel, composed of a nanoantenna transmitter and a protein receiver.  We assume that the nanoantenna is tuned to a specific frequency depending on the protein type. As such,  the interaction between the nanoantenna and the protein gives rise to a mechanical response~\cite{matellan2018no}. According to structural mechanics, if an external harmonic excitation has a frequency which matches one  of the natural frequencies of the system, then resonance occurs, and the vibrational amplitude increases~\cite{bassani2017terahertz}. This is the case with protein molecules as the value of their vibrational frequency is given as~\cite{carpinteri2017terahertz}
\begin{equation}
f_{protein}\approx\frac{1}{2\pi}\sqrt\frac{\kappa}{m}.
\end{equation}
 $\kappa$ and $m$ are the stiffness and the mass of the protein molecule, respectively. On average, proteins have  a stiffness of $10^2$ Nm$^{-1}$ and a mass of $10^{-24}$ kg yielding a vibrational frequency in the order of $10^{12}$, thereby matching the THz nanoantenna frequencies~\cite{jornet2013graphene}.

The capability to predict collective structural vibrational modes at THz frequencies has long attracted the research community. This interest has been fortified by  the development of THz spectroscopic techniques used to investigate the response of biomolecules~\cite{xie2014application}. In particular, vibrations can be dipole active, and thus probed using THz dielectric spectroscopy. The detected molecular motions in the picosecond range  correspond to collective vibrational modes or very fast conformational changes. An extensive review by Markelz explores measurements of the THz dielectric response on  molecules, where the author concludes that the response is highly sensitive to hydration, temperature, binding and conformational change~\cite{markelz2008terahertz}.

The investigated dielectric response of proteins includes both a  relaxational response  from the amino acid side chains along with a vibrational response from the correlated motions of the protein structure~\cite{knab2006hydration,son2014terahertz}. The authors in~\cite{carpinteri2017terahertz} associate such a vibrational phenomenon with the mechanical behavior of proteins, which act as oscillating structures in response to THz radiation. The induced electro-chemical force allows the identification of relevant resonant frequencies, which may enable a conceptual interpretation of the protein biological function. These frequencies, which range from hundreds of GHz  to tens of THz, can be mathematically captured using modal analysis. For instance, in lysozyme, a highly delocalized hinge-bending mode that opens and closes the binding cleft was found by normal mode calculations~\cite{brooks1985normal}.

In addition, measurements of chlorophyll proteins  showed an increase in the THz absorbance with denaturing, which arise due to the protein side chains' rotational motion~\cite{hua2007investigation}. Further, measurements reported in~\cite{turton2014terahertz} on lysozyme proteins showed sharp vibrational peaks at 1.15 and 2.80 THz. In addition, other measurements provided in~\cite{nicolai2016fingerprints}, showed that the Hsp70 protein, referred to as molecular chaperon, possessed distinct spectra for protein states at sub-THz frequencies. 
 
These measurements indicate that a nanoantenna can selectively target the vibrational mode of the protein related to either folding or unfolding and induce a conformational change. In fact, in~\cite{balu2008terahertz}, the authors provide a description of the modes of three proteins,  namely,  Rhodopsin,  Bacteriorhodopsin  and  D96N  bacteriorhodopsin  mutant.   This  gives an indication of the selectivity of these vibrational modes showcasing the capability to single out  proteins with a degree of accuracy. In addition to initiating information flow by inducing folding behavior, stimulating proteins by EM waves may provide knowledge of the misfolded protein structure. This potentially makes possible future efforts to rationally design drugs that prevent misfolding events along with the the evolution of certain conditions and diseases. 
 
\subsection{Boltzmann Distribution}

%
%
Signaling inside proteins results in a spring-like effect which shifts their minimum energy~\cite{orr2006mechanisms}. Protein structures are therefore investigated using energy functions where they obey statistical laws based on the Boltzmann distribution. On the one hand, the energy levels of  EM waves in the THz frequency band  are very low, corresponding to 1-12 meV~\cite{siegel2004terahertz,saeedkia2013handbook}. These values match  energies in the range  of $10^{-21}$ Joules. Since the energy expended $=$ force $\times$ distance, and we  deal with protein conformational changes, measured in nanometers~\cite{howard2001mechanics}, this will yield forces in the piconewton range. On the other hand, this energy scale conform  with energies  required for ATP hydrolysis,  ranging from $1$ $k_{b}T$ to $25$ $k_{b}T$ (here, $k_b$ is Boltzmann's constant and $T$ temperature in Kelvin ; 1 $k_{b}T$ at $300$ Kelvin $\approx$ $4 \times10^{-21}$)~\cite{howard2001mechanics}. Thereby, utilizing a THz force to drive a protein activity and a controlled molecular response is compatible with intra-body energetics.

The protein conformational change from one state to another  mimics a stretch activated channel. Based on statistical mechanics, the Boltzmann distribution provides probability that a system will be in a certain state as a function of the state's energy and system temperature. The probability of the protein existing in a certain state $i$ is
\begin{equation}
P_i=\frac{1}{Z} \exp \left[ \frac{-E_i}{k_bT} \right],
\label{eq:general1}
\end{equation}
where $E_i$ is the Gibbs free energy of the state and $Z$  is a  normalization factor which results from the constraint that the probabilities of all accessible states must add up to one, i.e., the normalization factor is given by
\begin{equation}
Z=\sum_{i=1}^{M}\exp \left[ \frac{-E_i}{k_bT} \right],
\label{eq:general2}
\end{equation}
where $M$ is the number of states accessible to the protein network. 

In our model, the Boltzmann distribution is altered to take into account the nanoantenna THz force. By applying an external force, $F$, the average position of the mechanotransduction channel is shifted, thereby impacting the state probability of the protein. This relation can be seen when finding the energy difference  between  states   given as
\begin{equation}
\Delta E=\Delta E^0_{ij}-F \Delta\ell,
\label{eq:energyf}
\end{equation}
where $\Delta E_{ij}^0= E_i-E_j $ is the difference in Gibbs free energy between initial state $i$ and final state $j$. $\Delta\ell$ denotes the change in the protein length, which corresponds to a conformational change in the protein structure requiring work $\phi(F) = F \Delta\ell$.  Gibbs free energy  expresses the thermodynamic energy reflecting the chemical potential between interacting proteins~\cite{rietman2016thermodynamic}. In fact, upon the change of  concentration of one molecular species, the reactions in which these molecular species participate are affected. Hence, a change in one protein concentration will percolate through the network changing its energy. The final result represents perturbation in the network leading to changes in the energetic landscape, or Gibbs energy of the molecule~\cite{rietman2017personalized}. If the protein is subject to a force, a natural reaction coordinate is the length of the protein in the direction of the force, and the total energy difference is given in~\eqref{eq:energyf}. 

\subsection{Stochastic Model of Protein Folding}
To model the stochasticity of  proteins involved  upon  triggering them by a THz force, we use the kinetic master equation  at the single protein level since it captures the chemical kinetics of the receptor~\cite{higham2008modeling}. Such approach is similar to the ones presented in~\cite{eckford2015information,eckford2016finite,eckford2018channel}. A transition rate matrix $R$ describes the rate at which a continuous time Markov chain moves between states. Elements $r_{ij}$ (for $i \neq j$) of matrix $R$ denote the rate departing from state $i$ and arriving in state $j$. Diagonal elements $r_{ii}$ are defined such that 
\begin{equation}
r_{ii}= \sum _{j\neq i} r_{ij}.
\end{equation}
In addition, the probability vector, $\mathbf{p}(t)$, as a function of time $t$ satisfies the transition rates via the differential equation
\begin{equation}
\frac{d\mathbf{p}(t)}{dt}=\mathbf{p}(t)R.
\label{eq:master_v2}
\end{equation}To represent the protein change of state as a discrete-time Markov chain, we discretize the time into steps of length $\Delta t$. As such, the  master equation provided in~\eqref{eq:master_v2} becomes 
\begin{equation}
\frac{d \mathbf{p}(t)}{dt}= \mathbf{p}(t) R = \frac{ \mathbf{p}(t+ \Delta t)- \mathbf{p}(t)}{\Delta t}+o(\Delta t).
\label{eq:discretization}
\end{equation}We neglect the terms of order $o(\Delta t)$ and manipulate~\eqref{eq:discretization} to have\begin{equation}
\begin{aligned}
 \mathbf{p}(t+ \Delta t) &= \Delta t \mathbf{p}(t)R+ \mathbf{p}(t)= \mathbf{p}(t)(I+ \Delta tR),
\label{eq:8}
\end{aligned}
\end{equation}
where $I$ is the identity matrix. If we denote $\mathbf{p}_{i}= \mathbf{p}(i\Delta t),$ we arrive at a discrete time approximation to~\eqref{eq:8} as,
\begin{equation}
 \mathbf{p}_{i+1}= \mathbf{p}_{i}(I+\Delta tR).
\end{equation}
Thus, we obtain a discrete-time Markov chain with a  transition probability matrix $Q$ given as
\begin{equation}
Q=I+\Delta t R.
\label{eq:matrixQ}
\end{equation}
\section{Protein Conformational Interaction  as a Communication System}  \label{Sec3}
We now discuss how induced protein interactions can be described as
information-theoretic communication systems: that is, in terms of input, output, and conditional input-output probability mass function (PMF).  The channel input is the nanoantenna force transmitted to the protein receptor: at the interface between the receptor and the environment, the receptor is sensitive to the induced force, undergoing changes in configuration  as  force 
is applied. The channel output is   the state of the
protein.  A Markov transition PMF dictates the input-output relationship  since the protein state depends on both the current input and the previous state. This relationship is given as  \begin{equation}
 p_{\mathbf{Y}|\mathbf{X}}(\mathbf{y}|\mathbf{x})=\prod_{i=1}^{n}\ p_{\mathbf Y_{i}| \mathbf X_{i},\mathbf Y_{i-1}}(y_{i}|x_{i},y_{i-1}),
\label{eq:cond1}
 \end{equation}where $p_{\mathbf Y_{i}|\mathbf X_{i}, \mathbf Y_{i-1}}(y_{i}|x_{i},y_{i-1})$ is provided according to the appropriate entry in matrix $Q$ given in~\eqref{eq:matrixQ} and $n$ is the fixed channel length.

For any communication system with inputs  $\mathbf{x}$ and outputs $\mathbf{y}$, the mutual information, $\mathcal{I}(\mathbf{X};\mathbf{Y})$,  provides the maximum information rate that may be transmitted reliably over the channel for a given input distribution. Maximizing this mutual information over the input distribution provides the channel capacity. This analysis is important in order for us to identify the maximum rate by which a protein can receive information and, thereby, we assess the impact of THz force on communication.  For tractability, we restrict inputs to the set of IID input distributions, where $p_{\mathbf{X}}(\mathbf{x})=\prod_{i=1}^{n}p_{\mathbf{X}}(x_i)$. The authors in~\cite{thomas2016capacity} showed that the IID input distribution was capacity achieving (i.e., max achievable rate) for two-state intensity-driven Markov chains. T he protein state $\mathbf{y}$ forms a time-homogeneous Markov chain given as
\begin{equation}
p_{\mathbf{Y}}(\mathbf{y})=\prod_{i=1}^{n} p_{\mathbf{Y}_{i}|\mathbf{Y}_{i-1}}(y_{i}|y_{i-1}), \label{eq:marg1}
\end{equation}
where $y_0$ is null and  
\begin{equation}
p_{\mathbf{Y}_{i}|\mathbf{Y}_{i-1}}(y_i|y_{i-1})=\sum_{x_{i}}p_{\mathbf{Y}_{i}|\mathbf{X}_{i},\mathbf{Y}_{i-1}}(y_i|x_i,y_{i-1})p_{\mathbf{X}}(x_{i}).
\label{eq:cond2}
\end{equation}
The mutual information can be written as
\begin{equation}
\begin{split}
\mathcal{I}(\mathbf{X};\mathbf{Y})=\sum_{i=1}^{n} \sum_{{y_i}} \sum_{{y_{i-1}}} \sum_{x_i} p_{\mathbf Y_i, \mathbf X_i,\mathbf Y_{i-1}}(y_i,x_i,y_{i-1})\\ \log\frac{p_{\mathbf Y_i| \mathbf X_i,\mathbf Y_{i-1}}(y_i|x_i,y_{i-1})}{p_{\mathbf Y_i|  \mathbf Y_{i-1}}(y_i|y_{i-1})}.
\label{eq:1}
\end{split}
\end{equation}
Thereafter, the channel capacity is given as
\begin{equation}
C= \max_{p_{\mathbf{X}}(\mathbf{x})} \,\ \mathcal{I}(\mathbf{X};\mathbf{Y}).
\end{equation} 
In our analysis, we deal with the input, $\mathbf{\mathbf{x}}$, as either a discrete or continuous  parameter. We use the bisection method to compute the capacity for the discrete case and deploy the Blahut-Arimoto (BA) algorithm to find the capacity for the continuous scenario. In fact, given an input-output transition matrix, the classical  BA algorithm is a general numerical method for computing the capacity channel~\cite{blahut1972computation}. The maximization of the mutual information is attained through an alternating maximization procedure to the global maximum. A variation of the BA algorithm is the constrained BA method, which incorporates an average power constraint on the channel inputs.

We provide several  capacity measures with different constraints for the EM-triggered protein communication channel. Specifically, we derive the capacity per channel use and with average energy constraint. Capacity per channel use is a suitable measure in applications involving targeted therapy or targeted drug delivery. The capacity with an average energy constraint is a useful  measure for  efficient intra-body communication, where both medium compatibility and safety metrics are practical constraints accounted for. In each case, the optimum input distribution and the resulting maximized capacity measures are attained.  
\section{Two-State Protein  Model}
\label{Sec4}
\begin{figure}
\centering
\includegraphics[width=0.2\textwidth]{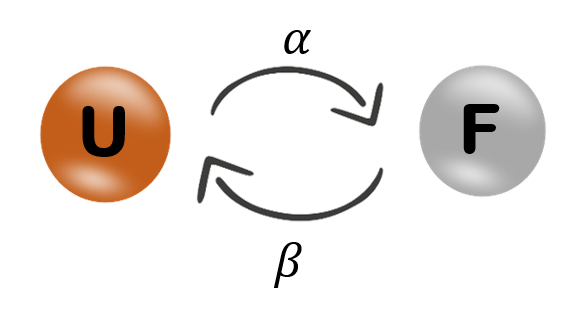}
\footnotesize
\caption{Two-state protein model represented by unfolded ($\mathbf{U}$) and folded ($\mathbf{F}$) Markov  states.}
\label{fig:model}
\end{figure}
\subsection{Mathematical Model}
In our two-state model, the protein resembles a binary biological switch, represented using  a  finite-state Markov chain. The states of the protein depicted are the folded, $\mathbf{F}$, and unfolded, $\mathbf{U}$, as  those govern the activation of biological  processes and chemical interactions. The input to our mechanotransduction channel is the force induced by the nanoantenna, while  the output is the state of the protein. In continuous time, the protein folding can be represented as a Poisson process, transitioning between $\mathbf{F}$ and  $\mathbf{U}$. We let $p_{\mathbf{Y}}(t)=[p_{\mathbf{F}}(t), p_{\mathbf{U}}(t)]$ denote the time-varying vector of state occupancy probabilities. 

As demonstrated in Fig.~\ref{fig:model}, in this system, the transition rate from unfolded, $\mathbf{U}$, to folded, $\mathbf{F}$, is $\alpha$, while the transition rate from $\mathbf{F}$ to $\mathbf{U}$ is $\beta$. The latter transition is considered  a relaxation process which returns the protein to the unfolded state. Such process is independent of the excitation signal since protein folding is entropically unfavorable~\cite{anfinsen1973principles}. The main reason for protein to get folded is to acquire its function. The function implies a general architecture of the protein which has to be stable  in time and flexible enough to allow the biological process to occur. Therefore native state of a protein is not necessarily the most stable one. To model the two-state  conformational change which captures the behavior of a protein, the normalization factor, provided in~\eqref{eq:general2}, is given by
\begin{equation}
Z=\exp\left[\frac{-E_{\mathbf{U}}}{k_{b}T}\right]+\exp\left[\frac{-E_\mathbf{F}}{k_{b}T}\right],
\label{eq:normz}
\end{equation}
where $E_{\mathbf{U}}$ and $E_{\mathbf{F}}$ denote the Gibbs free energies associated with the unfolding and folding states, respectively.
As such,  the steady-state  probability of the protein being in  one state, the folded for example, can be found from~\eqref{eq:general1} and~\eqref{eq:normz} as
\begin{equation}
p_{\mathbf{Y}}(y=\mathbf{F})=\frac{1}{1+\exp\left[ \frac{\Delta E}{k_{b}T} \right]}.
\label{eq:steady_state1}
\end{equation}
The transition rates controlling such two-state interaction are given by the rate matrix $R_{1}$ as \begin{equation}
R_{1}=\begin{bmatrix}-\alpha & \alpha \\
\beta & -\beta \\
\end{bmatrix}.
\end{equation}From~\eqref{eq:matrixQ}, the transition probability matrix yields
\begin{equation}
Q_{1}=\begin{bmatrix}1-\alpha \Delta t & \alpha\Delta t  \\
\beta \Delta t & 1-\beta\Delta t  
\label{eq:prob_matr}
\end{bmatrix}.
 \end{equation}
\subsection{Kinetic Detailed Balance}
The steady state probability is the eigenvector of the stochastic matrix, which can be found using the following relation 
\begin{equation}
\mathbf{p}_{\mathbf{Y}}(\mathbf{y})Q =\mathbf{p}_{\mathbf{Y}}(\mathbf{y}).
\label{eq:ss} 
\end{equation}
Hence, for our two-state Markov model the steady-states yield
\begin{equation}\label{eq:cases}
p_{\mathbf{Y}}(y)= 
\begin{cases}
    \frac{\alpha}{\alpha+\beta}, & y= \mathbf{F}\\
     \frac{\beta}{\alpha+\beta},& y= \mathbf{U}.
\end{cases}
\end{equation}
The relationship between $\alpha$  and $\beta$ can therefore be found by equating~\eqref{eq:steady_state1} and~\eqref{eq:cases} for $y= \mathbf{F}$, resulting in
\begin{equation}
{\beta}={\alpha}\, \exp\left( \frac{\Delta E}{k_{b}T} \right).
\label{eq:relationship_alpha_beta}
\end{equation}\eqref{eq:relationship_alpha_beta} satisfies the detailed balance theory, which has been formulated for kinetic systems~\cite{coester1951principle}.  Detailed balance ensures the compatibility of kinetic equations with the conditions for thermodynamic equilibrium. The rate constants pulling against an applied force resembles a biased random walk that allows the protein to perform work per unit step, i.e., $\phi(F)= F \Delta\ell$, in agreement with the second law of thermodynamics and as shown in~\eqref{eq:energyf}. 

Since the value of the energy, $\Delta E$, gets altered when the system is subject to an external force, the value of $\alpha$ (the probability of the forward transition rate) will also vary accordingly. As such, $\alpha$ can be divided into $\alpha_{\mathbf{NF}}$, the natural transition rate when no force is applied, and $\alpha_{\mathbf{AF}}$,  the transition rate when a force is applied, resulting in an average folding probability. The values of $\alpha_{\mathbf{NF}}$ and $\beta$ for different proteins can be found from experimental studies available  in the literature  since  protein folding  is a naturally occurring phenomenon driven by the change in Gibbs energy~\cite{fisher1999study}. Therefore,~\eqref{eq:relationship_alpha_beta} can take two different forms depending on whether the system is being subject to an external force or not as follows
\begin{numcases} {\beta=} 
    {\alpha_{\mathbf{NF}}}\, \exp\left( \frac{\Delta E}{k_{b}T}\right),\,\,\, \Delta E= \Delta E_{ij}^{0} \label{eq:relationship_alpha_beta1} \\
     \alpha_{\mathbf{AF}}\,\exp\left( \frac{\Delta E}{k_{b}T}\right),\,\,\, \Delta E= \Delta E_{ij}^{0}+ \phi(F)  \label{eq:relationship_alpha_beta2}
\end{numcases}
Here, $\mathbf{NF}$ and $\mathbf{AF}$ correspond to No Force and Applied Force, respectively.
\subsection{Capacity  of Two-State Protein  Conformation}
\subsubsection{Discrete Case}
   Based on our developed model, we let $\mathbf{\mathbf{x}}$ denote a binary  input  which stimulates the protein. This input is induced either due to intra-body interactions with no external force or could be triggered due to an applied THz\ nanoantenna force, in which  $\mathbf{\mathbf{x}}\in \left \{ \mathbf{NF}, \mathbf{AF}\right\}$.  The channel output is the state of the protein given as either  unfolded or folded, where $\mathbf{y}\in \left\{ \mathbf{U}, \mathbf{F}\right\}$.  We have, as a result, a discrete channel, where the inputs and outputs form vectors. In order to find the capacity, we follow the formulation presented in Sec. III. Assuming the previous state of the protein, $y_{i-1}=\mathbf{U}$, 
 we have\begin{equation}
 \begin{split}
p_{\mathbf{Y}_{i}|\mathbf{Y}_{i-1}}(\mathbf{F}|\mathbf{U})&=\sum_{x_{i}}p_{\mathbf{Y}_{i}|\mathbf{X}_{i},\mathbf{Y}_{i-1}}(\mathbf{F}|x_i,\mathbf{U})p_{\mathbf{X}}(x_{i})\\
&=p_{\mathbf{NF}}\alpha_{\mathbf{N}\mathbf{F}}+p_{\mathbf{AF}}\alpha_{\mathbf{A}\mathbf{F}}=\bar \alpha, 
\label{eq:alpha_bar}
\end{split}
\end{equation}and  $p_{\mathbf{Y}_{i}|\mathbf{Y}_{i-1}}(\mathbf{U}|\mathbf{U})=1-\bar \alpha$. Here, $\bar \alpha$ represents the average folding probability. 
On the other hand, if $y_{i-1}=\mathbf{F}$,
\begin{equation}
\begin{split}
p_{\mathbf{Y}_{i}|\mathbf{Y}_{i-1}}(\mathbf{U}|\mathbf{F})&=\sum_{x_{i}}p_{\mathbf{Y}_{i}|\mathbf{X}_{i},\mathbf{Y}_{i-1}}(\mathbf{U}|x_i,\mathbf{F})p_{\mathbf{X}}(x_{i})
\\
&=\beta,
\end{split}
\end{equation}
and $p_{\mathbf{Y}_{i}|\mathbf{Y}_{i-1}}(\mathbf{F}|\mathbf{F})=1-\beta$. The transition probability matrix provided in~\eqref{eq:prob_matr} can now be written as
\begin{equation} \label{eq:sys_mat} 
\bar Q_{1}=\begin{bmatrix}1-\bar\alpha\Delta t & \bar\alpha\Delta t  \\
\beta \Delta t & 1-\beta\Delta t  \\
\end{bmatrix}.  
 \end{equation}
 In addition, the steady state probabilities given in~\eqref{eq:cases} are adjusted to take into account the average folding probability, $\bar\alpha$.
 
 The mutual information, $\mathcal{I}(\mathbf{X};\mathbf{Y})$, which was given in~\eqref{eq:1}, can also be represented as
\begin{equation}
\mathcal{I}(\mathbf{X};\mathbf{Y})=H(Y_i|Y_{i-1})-H(Y_i|X_i,Y_{i-1}), 
\label{eq:mutual_info3}
\end{equation}
for $i\in \{1,2,...,n\}$. To compute~\eqref{eq:mutual_info3}, we use the binary entropy function as follows
\begin{equation}
\mathcal{H} (p)=-p\log p-(1-p)\log (1-p).
\end{equation}
Then, each term in the right hand side of~\eqref{eq:mutual_info3}, is dealt with separately. $H(Y_i|Y_{i-1})$ yields \begin{equation}
\begin{split}
&=p_{\mathbf{Y}}({\mathbf{U}})H(Y_i|Y_{i-1}=\mathbf{U})+p_{\mathbf{Y}}({\mathbf{F}})H(Y_i|Y_{i-1}=\mathbf{F})\\ &=\frac{\beta}{\bar\alpha+ \beta} \mathcal H(\bar\alpha) +\frac{\bar\alpha}{\bar\alpha+ \beta}\mathcal H(\beta).
\end{split}
\end{equation}
In a similar manner, $H(Y_i|X_i,Y_{i-1})$ results in
\begin{equation}
\begin{split}
&=\sum_{x_i}p_\mathbf{X}(x_i)p_{\mathbf{Y}}(\mathbf{U})H(Y_i|X_i=x_i,Y_{i-1}=\mathbf{U})
\\ &+\sum_{x_i}p_\mathbf{X}(x_i)p_{\mathbf{Y}}(\mathbf{F})H(Y_i|X_i=x_i,Y_{i-1}=\mathbf{F})\\
 &=\frac{\beta}{\bar\alpha+ \beta} \left( p_{\mathbf{NF}}\mathcal{H}(\alpha_{\mathbf{N}\mathbf{F}})+p_{\mathbf{AF}}\mathcal{H}(\alpha_{\mathbf{A}\mathbf{F}}) \right)+\frac{\bar\alpha}{\bar\alpha+ \beta}\mathcal{H}(\beta).
\end{split}
\end{equation}
By substituting back into~\eqref{eq:mutual_info3}, the mutual information yields
\begin{equation}
\begin{aligned} 
\mathcal{I}(\mathbf{X};\mathbf{Y})= \frac{\beta}{\bar\alpha+ \beta} \left( \mathcal H(\bar\alpha)-p_{\mathbf{NF}}\mathcal{H}(\alpha_{\mathbf{N}\mathbf{F}})- p_{\mathbf{AF}}\mathcal{H}(\alpha_{\mathbf{A}\mathbf{F}})\right).\\ \label{eq:final_eq}
=\frac{\mathcal{H}(p_{\mathbf{NF}}\alpha_{\mathbf{N}\mathbf{F}}+p_{\mathbf{AF}}\alpha_{\mathbf{A}\mathbf{F}})-p_{\mathbf{NF}}\mathcal{H}(\alpha_{\mathbf{N}\mathbf{F}})- p_{\mathbf{AF}}\mathcal{H}(\alpha_{\mathbf{A}\mathbf{F}})}{1+\left( p_{\mathbf{NF}}\alpha_{\mathbf{N}\mathbf{F}}+p_{\mathbf{AF}}\alpha_{\mathbf{A}\mathbf{F}} \right)/\beta}. 
\end{aligned}
\end{equation}
Finally, the  capacity of the two-state model is found by maximizing~\eqref{eq:final_eq} with respect to the nanoantenna applied force as 
\begin{multline}
C=\max _{p_\mathbf{AF}} \frac{\mathcal{H}(p_{\mathbf{NF}}\alpha_{\mathbf{N}\mathbf{F}}+p_{\mathbf{AF}}\alpha_{\mathbf{A}\mathbf{F}})}{1+\left( p_{\mathbf{NF}}\alpha_{\mathbf{N}\mathbf{F}}+p_{\mathbf{AF}}\alpha_{\mathbf{A}\mathbf{F}} \right)/\beta}\\
+\frac{-p_{\mathbf{NF}}\mathcal{H}(\alpha_{\mathbf{N}\mathbf{F}})- p_{\mathbf{AF}}\mathcal{H}(\alpha_{\mathbf{A}\mathbf{F}})}{1+\left( p_{\mathbf{NF}}\alpha_{\mathbf{N}\mathbf{F}}+p_{\mathbf{AF}}\alpha_{\mathbf{A}\mathbf{F}} \right)/\beta}
\label{eq:capacity}.
\end{multline}
It is sufficient to maximize over $p_{\mathbf{AF}}$ since $p_{\mathbf{NF}}=1-p_{\mathbf{A}\mathbf{F}}$. \\
\subsubsection{Continuous Case}
In the previous part,  we developed the model as a discrete case given a binary input binary output system. Nonetheless, an in-depth picture for the capacity associated with protein conformational transitions is attained by applying a continuous input. By having the nanoantenna force transmit continuously, the capacity versus applied force can be studied over a range of values. This is achieved by expanding $\bar\alpha$ in~\eqref{eq:alpha_bar} to become 
\begin{equation}
\bar \alpha= \alpha_{\mathbf{N}\mathbf{F}}p_{\mathbf{N}\mathbf{F}}+\sum_{i=1}^{N-1}  \alpha_{\mathbf{A}\mathbf{F}}(f_i)p_{\mathbf{A}\mathbf{F}}(f_i), \,\,\, \label{eq:alphabar}
\end{equation}
 where $p_{\mathbf{AF}}(f_i)$ denotes the probability of applying a force, $f_i$, towards the protein.  The dependency of $\alpha_{\mathbf{AF}}$ on the force factor has been demonstrated in~\eqref{eq:relationship_alpha_beta2}.  

We find the capacity for the two-state model under the constraint of a maximum applied
force per channel use as
\begin{equation}
\begin{aligned}
& \underset{p_{\mathbf{A}\mathbf{F}}}{\text{max} \,\,}
& & \mathrm{\mathcal{I}(\mathbf{X};\mathbf{Y})} \\
& \text{subject to}
& & 0\leq F_{applied}\leq  \ F_{{max}}. \\
\end{aligned}
 \label{eq:sys_const1}
\end{equation}
${F}_{{max}}$ in this case is the maximum amount of nanoantenna applied force and ${p_\mathbf{AF}}$ is the probability vector of applied forces.
The objective function in~\eqref{eq:sys_const1} is concave with respect to the input probability vector and the constraint is linear; hence, the  optimization problem is concave. Therefore, the solution of the problem can be obtained using  the BA algorithm. The algorithm begins with the transition probability matrix, initially defined in~\eqref{eq:sys_mat}, but extended to take into account the $N$ maximum force samples along with an arbitrary but valid, choice for ${p_\mathbf{AF}}$. Since the mutual information in~\eqref{eq:sys_const1} is concave in terms of the input probability, the output of the algorithm is the optimal, capacity-achieving, input probability distribution, ${\hat p_\mathbf{AF}}$.

\section{Multi-State Protein Model}
\label{Sec5}
\subsection{Mathematical  Model}
Successive events occur inside a living cell through  a sequence of  protein activation in which signaling cascades are often illustrated by kinetic schemes.  Although a node in a network is represented by a single  protein, the protein itself can have multiple gene products with many conformations. Each node of the  protein can  slightly differ in sequence. Such differences allow a node to bind with hundreds of partners at different times and perform many essential biological functions~\cite{tsai1996protein}. 

In this section, we further extend the two-state protein conformation model to consider the transition between different protein configurations in order to more accurately resemble the protein signaling pathway especially when there are multiple folding routes from different starting points~\cite{graves1999protein}. As such, we generalize the two-state model presented previously to take into account multiple-states. The selectivity attained by using THz signals allows us to target specific links in a given network in order to create controlled interactions. These macroscopic interactions resemble the creation or removal of edges between nodes in a graph~\cite{vishveshwara2002protein}. By targeting the THz force on specific locations of the protein molecule, distinct responses can be induced.

We let $\mathbf{p}_{\mathbf{Y}}(t)=\left[ p_{y_{1}}(t), p_{{y_{2}}}(t),...., p_{y_{m+1}}(t) \right]$ be the probability vector accounting for $n=m+1$ states and $m$ links. In this case, the generalized rate matrix yields
\begin{equation}
R=\begin{bmatrix}-\alpha_1 & \alpha_1 & 0 & 0 & ....& ....  \\
\beta_{1}\ & -(\beta_1+\alpha_2) & \alpha_2 & 0&....&.... \\
0 & \beta_2 & -(\beta_2+\alpha_3) & \alpha_3&....&.... \\
: & : & : & :&:&:\\
: & : & : & :&\beta_m& -\beta_m\\
\end{bmatrix}.
\end{equation}Following the same formulation presented in~\eqref{eq:matrixQ}, the generalized probability matrix is  given in~\eqref{eq:prob_matrixg}. We note that throughout the analysis, we will use $\bar Q$ rather than $Q$, where each $\alpha_{j}$ is replaced by $\bar \alpha_j$, indicating an average state change probability.
\begin{figure*}
\centering
\begin{minipage}{0.75\textwidth}
\begin{align}
\bar Q=\left[ \begin{array}{ccccccc}
1-\bar\alpha_1\Delta t & \bar\alpha_1 \Delta t & 0 & 0&...& ...   \\
\beta_{1} \Delta t\ &  1-(\beta_1+\bar\alpha_2)\Delta t & \bar\alpha_2 \Delta t & 0& ... &...  \\
0 & \beta_2 \Delta t & 1-(\beta_2+\bar\alpha_3)\Delta t & \bar\alpha_3 \Delta t & ...& ... \\
: & : & : & : & : & :  \\
: & : & : & : & \beta_m \Delta t \ & 1-\beta_m \Delta t\\
\end{array} \right].
\label{eq:prob_matrixg}
\end{align}
\hrule
\end{minipage}
\end{figure*}

To compute the mutual information, $\mathcal{I}(\mathbf{X};\mathbf{Y})$, for the multi-state conformational model, we follow the same approach as in the previous section, where we provide a generalization of the formulation. First, following~\eqref{eq:mutual_info3}, we first compute $H(Y_i|Y_{i-1})$ as \begin{equation}
\begin{split}
&=p_{\mathbf{Y}}(y_1)\mathcal{H}(\bar\alpha_{1})+\sum_{j=2}^{m} p_{\mathbf{Y}}(y_j)\bigg( \mathcal{H}(\beta_{j-1})+\mathcal{H}(\bar\alpha_{j}) \bigg)\\
&\hspace*{0.5in} +
p_{\mathbf{Y}}(y_{m+1})\mathcal{H}(\beta_{m}). 
\end{split}
\end{equation}
Then, we find $H(Y_i|X_i,Y_{i-1})$ as
\begin{equation}
\begin{split}
 &=p_{\mathbf{Y}}(y_1)\bigg(p_{\mathbf{AF_{1}}}\mathcal{H}(\alpha_{\mathbf{AF_1}})+p_{\mathbf{NF_{1}}}\mathcal{H}(\alpha_{\mathbf{NF_1}})\bigg)  \\ &+ \sum_{j=2}^{m} p_{\mathbf{Y}}(y_j)\bigg(\mathcal{H} (\beta_{j-1})+\bigg( p_{\mathbf{AF_{j}}}\mathcal{H}(\alpha_{\mathbf{AF_j}})+p_{\mathbf{NF_{j}}}\mathcal{H}(\alpha_{\mathbf{NF_j}})\bigg)\bigg)\\
 & \hspace*{0.5in} + p_{\mathbf{Y}}(y_{m+1})\mathcal{H}(\beta_{m}).
\end{split}
\end{equation}
Substituting back in~\eqref{eq:mutual_info3} we get
\begin{multline}
\mathcal{I}(\mathbf{X};\mathbf{Y}) = \sum_{j=1}^{m}p_{\mathbf{Y}}(y_j)\mathcal{H}(\bar\alpha_{{j}})- \sum_{j=1}^{m} p_{\mathbf{Y}}(y_j)\\ \bigg( p_{\mathbf{AF}_{{j}}}\mathcal{H}(\alpha_{\mathbf{AF}_{j}})+p_{\mathbf{NF}_{j}}\mathcal{H}(\alpha_{\mathbf{NF}_{j}})\bigg).
\label{eq:final_eqg}
\end{multline}The capacity of the multi-state protein model is found by maximizing~\eqref{eq:final_eqg} with respect to the nanoantenna applied force as 
\begin{multline}
C=\max _{p_\mathbf{AF}} \bigg[ \sum_{j=1}^{m}  p_{\mathbf{Y}}(y_j)\mathcal{H}(\bar\alpha_{j}) - \sum_{j=1}^{m} p_{\mathbf{Y}}(y_j)\\\bigg( p_{\mathbf{AF}_{{j}}}\mathcal{H}(\alpha_{\mathbf{AF}_{j}})+p_{\mathbf{NF}_{j}}\mathcal{H}(\alpha_{\mathbf{NF}_{j}}\bigg)\bigg].
\label{eq:capacityg}
\end{multline}
In this case, $p_{\mathbf{AF}}$ is a vector constituting the probability of force applied to the $m$ links. 

\subsection{Example: Four State Protein  Model}
\begin{figure}[h]
\centering
\includegraphics[width=0.46\textwidth]{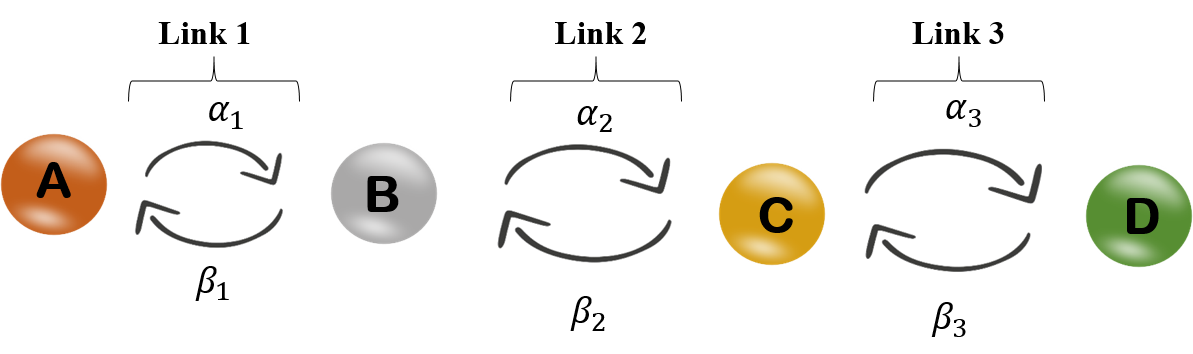}
\footnotesize
\caption{Multi-state protein model with several transitions.}
\label{fig:model_2}
\end{figure}
To show the applicability of the protein multi-state model, we apply it to a 4 state protein chain. We have the probability occupancy vector as, $\mathbf{p}(t)=\left[p_{\mathbf{A}}(t), p_{\mathbf{B}}(t), p_{\mathbf{C}}(t), p_{\mathbf{D}}(t)\right].$ The relationship between the states is formulated using a Markov transition PMF, which is previously given in~\eqref{eq:cond1}  and~\eqref{eq:cond2}. Hence,  based on Fig.~\ref{fig:model_2},
if the previous state, $y_{i-1}=\mathbf{A}$,
 we have\begin{equation}
 \begin{split}
p_{\mathbf{Y}_{i}|\mathbf{Y}_{i-1}}(\mathbf{B}|\mathbf{A})&=\sum_{x_{i}}p_{\mathbf{Y}_{i}|\mathbf{X}_{i},\mathbf{Y}_{i-1}}(\mathbf{B}|x_i,\mathbf{A})p_{\mathbf{X}}(x_{i})\\
&=p_{\mathbf{NF_{1}}}\alpha_{\mathbf{N}\mathbf{F_{1}}}+p_{\mathbf{AF_{1}}}\alpha_{\mathbf{A}\mathbf{F_1}}=\bar \alpha_{1}, 
\end{split}
\end{equation}
and  $p_{\mathbf{Y}_{i}|\mathbf{Y}_{i-1}}(\mathbf{A}|\mathbf{A})=1-\bar \alpha_{1}$. On the other hand, if $y_{i-1}=\mathbf{B}$,
\begin{equation}
\begin{split}
p_{\mathbf{Y}_{i}|\mathbf{Y}_{i-1}}(\mathbf{A}|\mathbf{B})&=\sum_{x_{i}}p_{\mathbf{Y}_{i}|\mathbf{X}_{i},\mathbf{Y}_{i-1}}(\mathbf{A}|x_i,\mathbf{B})p_{\mathbf{X}}(x_{i})
\\
&=\beta_1,
\end{split}
\end{equation}
and $p_{\mathbf{Y}_{i}|\mathbf{Y}_{i-1}}(\mathbf{B}|\mathbf{B})=1-(\beta_1 +\bar\alpha_2$). The relationship between the remaining states follows accordingly.

Using~\eqref{eq:ss}, the steady state probabilities are found as \begin{equation}\label{eq:cases2}
p_{\mathbf{Y}}(y)= 
\begin{cases}
\frac{\beta_1 \beta_2 \beta_3}{\beta_1\beta_2 \beta_3+\bar\alpha_1\beta_2 \beta_3+  \bar\alpha_1\bar\alpha_2\beta_3+\bar\alpha_1\bar\alpha_2\bar\alpha_3}, & y= \mathbf{A}\\ \\
     \frac{\bar\alpha_1 \beta_2 \beta_3}{\beta_1\beta_2 \beta_3+\bar\alpha_1\beta_2 \beta_3+  \bar\alpha_1\bar\alpha_2\beta_3+\bar\alpha_1\bar\alpha_2 \bar\alpha_3},& y= \mathbf{B} \\ \\
     \frac{\bar\alpha_1 \bar\alpha_2 \beta_3}{\beta_1\beta_2 \beta_3+\bar\alpha_{1}\beta_2 \beta_3+ \bar\alpha_1\bar\alpha_2\beta_3+\bar\alpha_2\bar\alpha_3\bar\alpha_1},& y= \mathbf{C}\\ \\
     \frac{\bar\alpha_1 \bar\alpha_2 \bar\alpha_3 }{\beta_1\beta_2 \beta_3+\bar\alpha_1\beta_2 \beta_3+ \bar\alpha_1\bar\alpha_2\beta_3+\bar\alpha_1\bar\alpha_2\bar\alpha_3},& y= \mathbf{D}
\end{cases}
\end{equation}
In~\eqref{eq:cases2}, we have considered the steady states after a force has been applied to the system, i.e., each $\alpha_{j}$ is replaced by $\bar\alpha_{j}$. We note also that the same relationship between $\alpha$ and $\beta$ holds as~\eqref{eq:relationship_alpha_beta} in Sec.~\ref{Sec3}. Finally, both the mutual information and capacity are found by substituting the given states in~\eqref{eq:final_eqg} and~\eqref{eq:capacityg} accordingly.

\subsection{Capacity with Average Energy Constraint} 
A variation on the optimization in~\eqref{eq:sys_const1} is when the average energy of applied nanoantenna force per channel use is also constrained. In this case, the constrained BA algorithm is deployed to find the capacity of the multi-state protein model. The resulting optimization problem is given as
\begin{equation}
\begin{aligned}
& \underset{p_{\mathbf{A}\mathbf{F}}}{\text{max} \,\,}
& & \mathrm{\mathcal{I}(\mathbf{X};\mathbf{Y})} \\
& \text{subject to}
& & \sum_{i} p_{AF_i} E_{i} \leqslant E^{max}, \\
&&& 0\leq p_{AF_{i}}\leq 1.   \\
\end{aligned}
 \label{eq:sys_const2}
\end{equation}
$E_i$ is the energy applied to  link $i$. The capacity with average energy constraint $E^{max}$ is defined as  
\begin{eqnarray}
C & = &\max_{p_\mathbf{{AF}}}\left[ \sum_{i}   p_{AF_{i}} \bar Q\log \frac{\bar Q}{\sum_{i}p_{AF_i}\bar Q} \right. \nonumber \\  & & \hspace*{0.65in} \left. - \lambda(\sum_{i}p_{AF_{i}}E_{i}-E^{max})\right].
\label{eq:const2}
\end{eqnarray}
Here, $\bar Q$ is the transition probability matrix defined in~\eqref{eq:prob_matrixg}.
The cost function in~\eqref{eq:const2} is parametrized using Lagrange multiplier $\lambda$. The procedure followed to optimize the input distribution is similar to that without the average energy constraint. The additional step involves obtaining a value for $\lambda$ after updating the distribution vector $p_{\mathbf{AF}}$. This can be obtained using a simple bisection search.

\section{Numerical Results}
\label{Sec7}
In this section,  we demonstrate the results of numerically simulating our developed models. The  aim of the presented work is to find the information rates by which  protein molecules convey information  when triggered by THz nanoantennas.  Several scenarios are presented to take into account different protein configurations undergoing either single or multiple signaling interactions.
\subsection{Discrete Case Result}
  In our discrete scenario, the system is binary, where the nanoantenna force is either present or absent as mathematically formulated in Sec.~\ref{Sec4}. The mutual information is calculated from the analytically derived model and the capacity is computed using a bisection search. This method is guaranteed to converge to a root, which is the value of $p_{\mathbf{AF}}$ that maximizes the capacity in our case. The discrete scenario  proves the existence of a communication channel, where information can be transmitted upon triggering the protein by THz EM waves. 

\begin{figure}[htp]
\subfigure[]{%
  \includegraphics[height=5.3 cm, width=7cm]{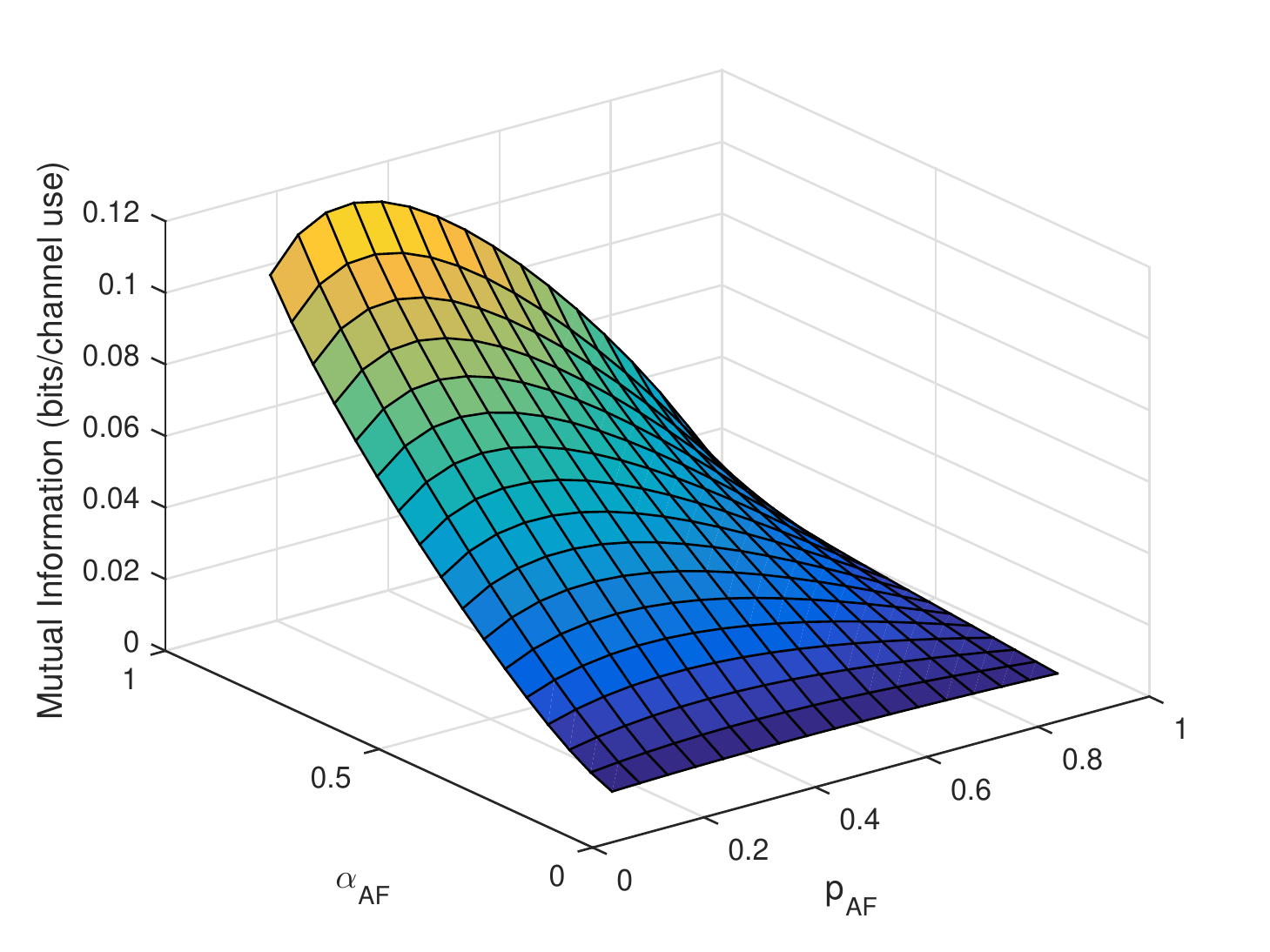}%
}
\subfigure[]{%
  \includegraphics[height=5.3 cm, width=7cm]{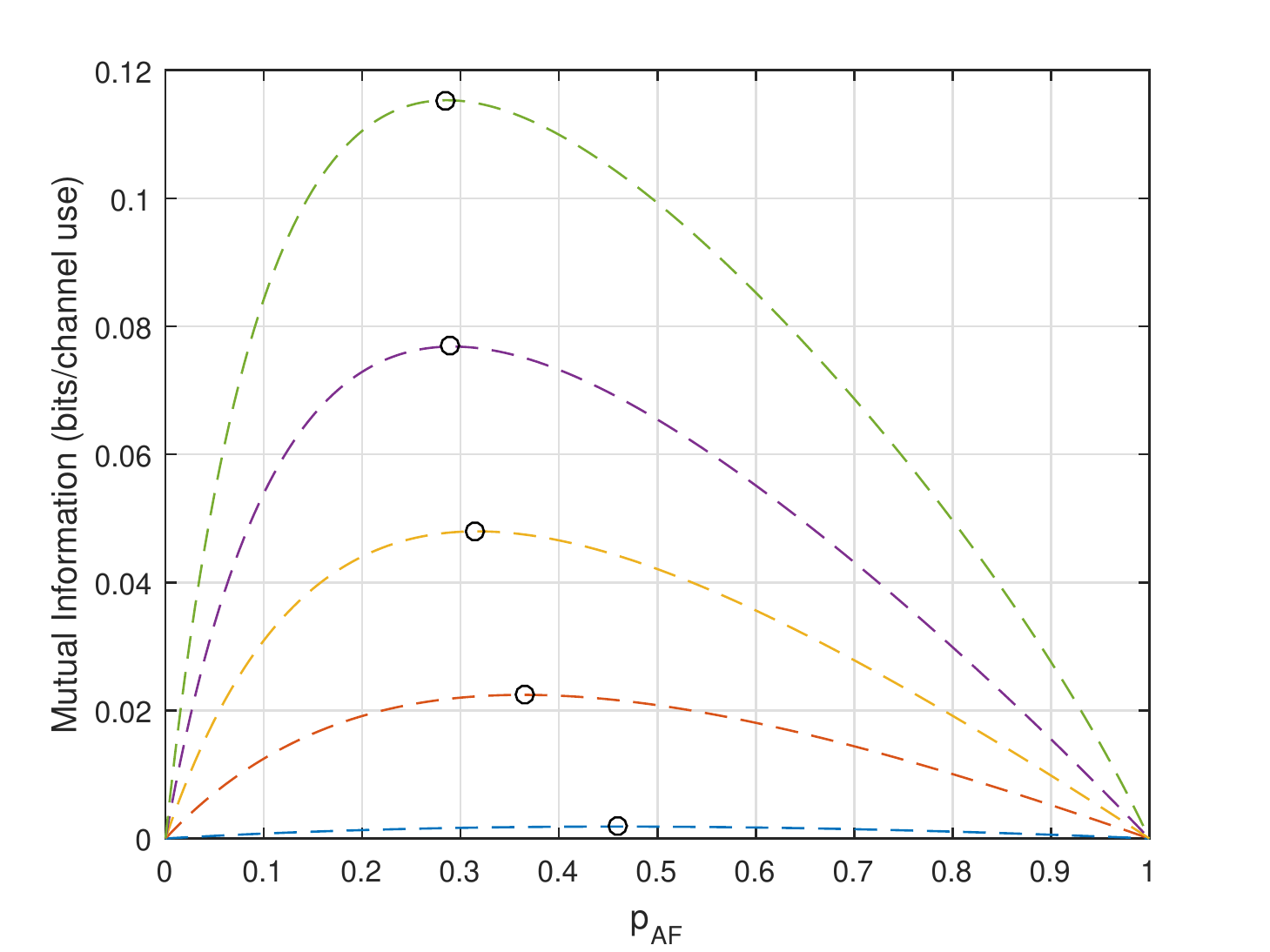}%
}

\caption{(a) 3D contour plot of the mutual information curve where $p_{\mathbf{AF}}$ and $\alpha_{\mathbf{AF}}$ are varied. (b) 2D plot showing the maximizing values of $p_{\mathbf{AF}}$ by circles. $\alpha_{\mathbf{NF}}=0.1$ and $\beta=0.1$, while $\alpha_{\mathbf{AF}}$ varies from the bottom from $0.1$ to $0.9$ with a $0.2$ increment.}
\label{fig:combined1}
\end{figure}

\begin{figure}[htp]
\centering
\subfigure[]{%
  \includegraphics[height=5.3 cm, width=7cm]{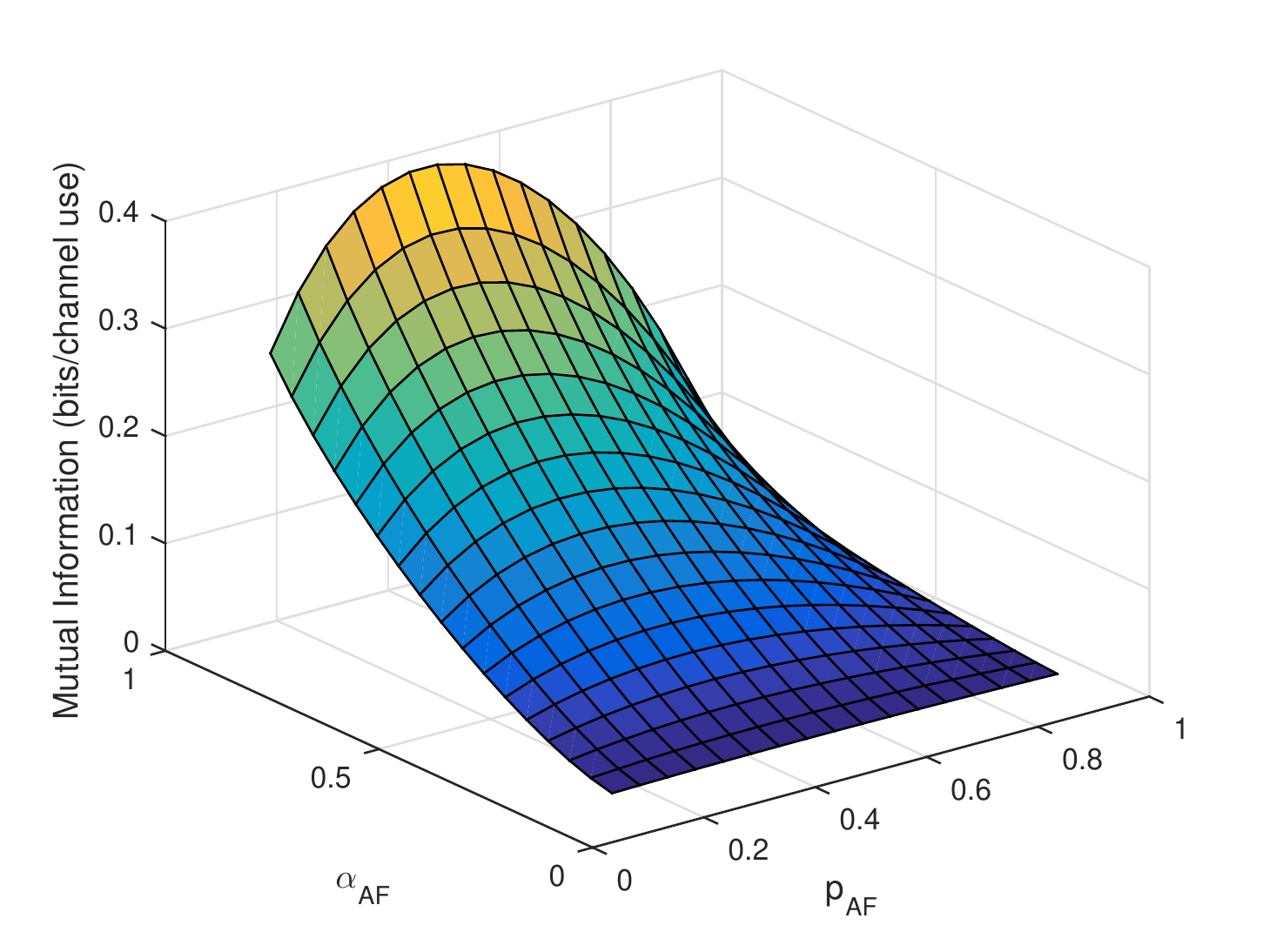}%
}

\subfigure[]{%
  \includegraphics[height=5.3 cm, width=7cm]{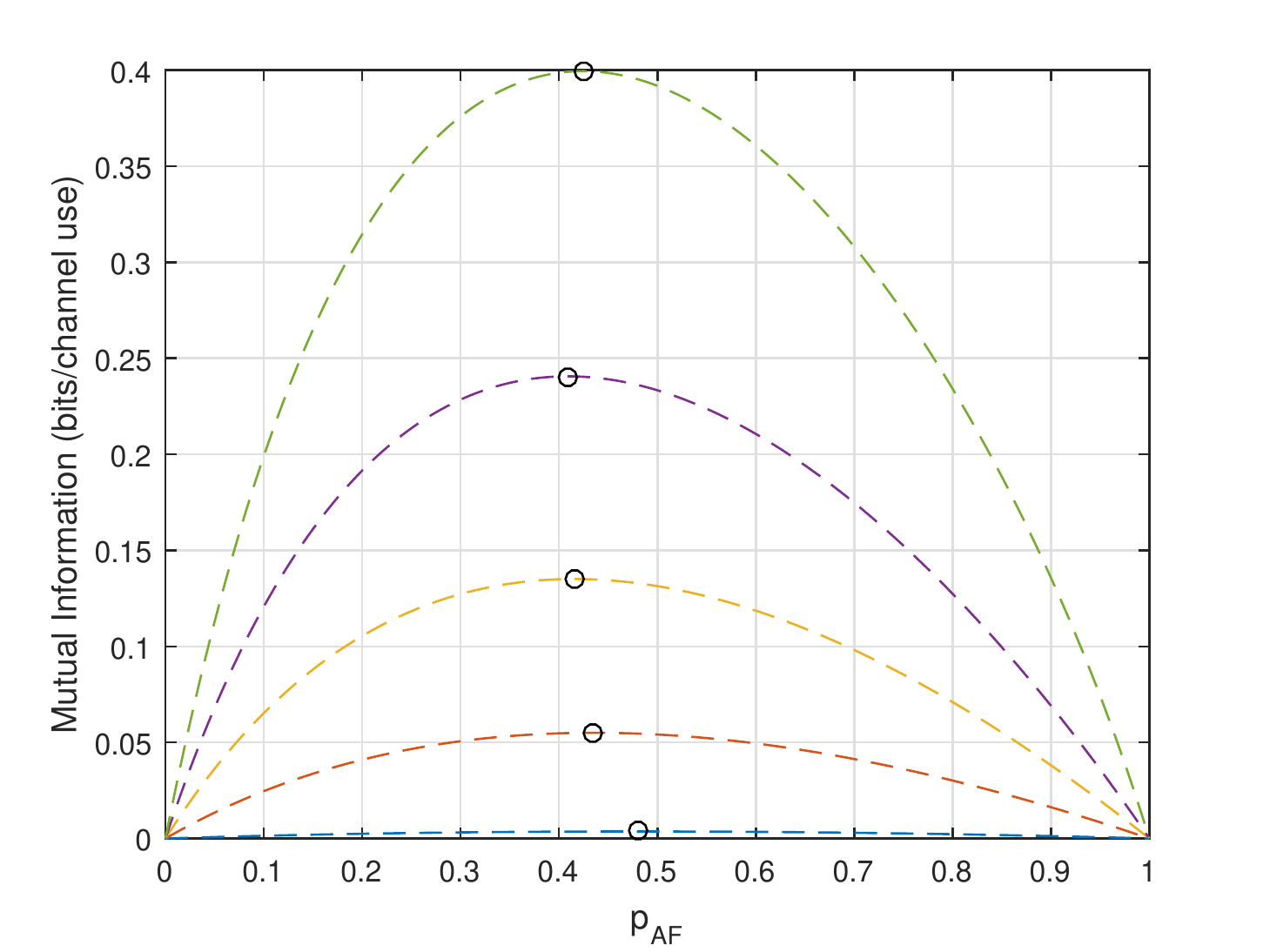}%
}

\caption{(a) 3D contour plot of the mutual information curve where  $p_{\mathbf{AF}}$ and $\alpha_{\mathbf{AF}}$ are varied. (b) 2D plot showing the maximizing values of $p_{\mathbf{AF}}$ by circles.  $\alpha_{\mathbf{NF}}=0.1$ and $\beta=0.9$, while $\alpha_{\mathbf{AF}}$ varies  from the bottom from $0.1$ to $0.9$ with  $0.2$ increment.}
\label{fig:combined2}
\end{figure}

Figs.~\ref{fig:combined1} and~\ref{fig:combined2} illustrate the mutual information curves  for $\beta=0.1$ and $\beta=0.9$, respectively. The value of $\alpha_{\mathbf{NF}}$ is fixed to $0.1$ while the values of $\alpha_{\mathbf{AF}}$ vary for both cases. As expected, the higher the value of $\alpha_{\mathbf{AF}}$, the higher the capacity since the value of $\alpha_{\mathbf{AF}}$ corresponds to the probability of folding. In addition, we notice that higher values of $\beta$ indicate a higher capacity. This observation can be deduced from~\eqref{eq:final_eq},  where an increased value of $\beta$ corresponds to a higher value of $\mathcal{I}(\mathbf{X};\mathbf{Y})$. The values of $p_{\mathbf{AF}}$ which maximize the capacity are clearly indicated using circles on the demonstrated 2D plots of the mutual information curves.

\subsection{Capacity Per Channel Use Result}
For the case of a continuous force, the BA algorithm is deployed to find the capacity. The attained result further fortifies the discrete case  by providing a more detailed analysis of how the capacity varies as a function of force. We utilize the relationships given in~\eqref{eq:alphabar} and~\eqref{eq:sys_const1} to simulate this scenario. Protein conformational changes are measured in nanometers (nm) and forces are given on  the scale of piconewtons (pN)~\cite{valle2017multidomain}. The value for the protein conformational distance was fixed at $\Delta\ell= 2$ nm for maximum forces ranging between $0-100~$pN. The selected force range of the nanoantenna reflects THz transmissions based on intra-body link budget analysis~\cite{7955066} and force sensitivity at the cellular level~\cite{matellan2018no}.

\begin{figure}[!h]
\centering
\includegraphics[height=5.25cm, width=8.1cm]{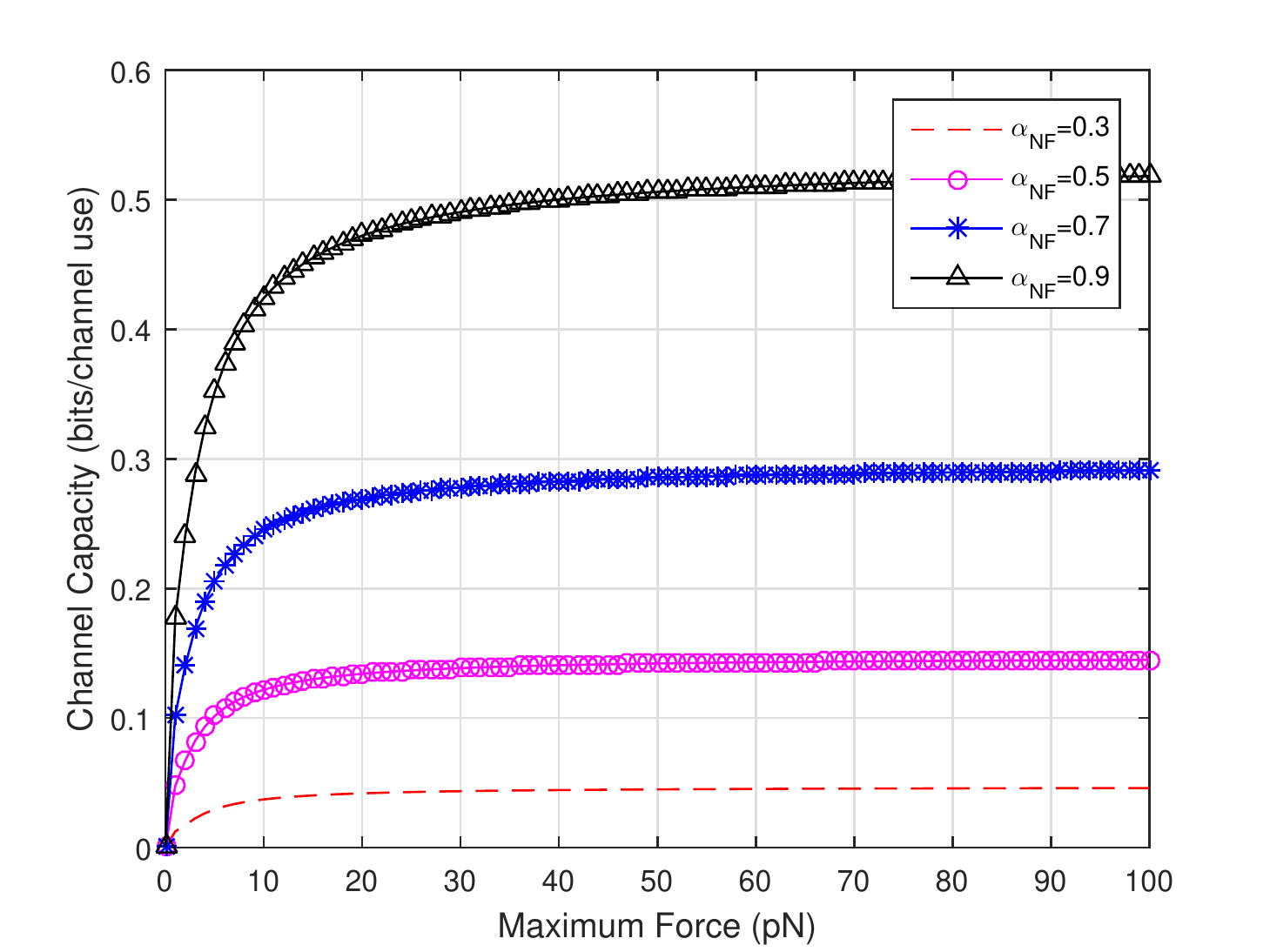}
\footnotesize
\caption{The channel capacity  as a function of the nanoantenna applied force. The value of $\beta$ is fixed to $0.9$ while the value of $\alpha_{\mathbf{NF}}$ varies.}
 \label{fig:cont1}
\end{figure}

Fig.~\ref{fig:cont1} demonstrates the capacity as a function of the applied nanoantenna force. We observe that given a fixed  value of $\beta$ and  $\alpha_{\mathbf{NF}}$, the value of the capacity increases upon increasing the nanoantenna applied force. In addition, the higher the value of $\alpha_{\mathbf{NF}}$, the higher the achieved capacity for the value of $\beta=0.9$. In order to understand such behavior, the change in Gibbs free energy, $\Delta E_{ij}^0$, must be examined. In fact, $\Delta E_{ij}^0$ is computed using the relationship presented in~\eqref{eq:relationship_alpha_beta1}, which is rearranged to yield
\begin{equation}
\Delta E_{ij}^0=k_{b}T \ln\left[\frac{\alpha_{\mathbf{NF}}}{\beta}\right].
\label{eq:concluded_relation}
\end{equation}
By increasing the value of $\alpha_{\mathbf{NF}}$, $\Delta  E_{ij}^{0}$ witnesses increments until it approaches equilibrium ($\Delta E_{ij}^{0}=0$) at $\alpha_{\mathbf{NF}}=0.9$. The equilibrium state indicates a chemical balance, where  no work should be done on the system as it is currently in a stable state. As such, the amount of force directed from the nanoantenna will be  solely dedicated  towards  increasing the capacity at which the protein receives information. Hence, no force  will be lost in order to first stabilize the system and then contribute to the capacity. Even for low values of $\alpha_{\mathbf{NF}}$, a capacity-achieving channel is attained upon applying a force. This indicates that the presented EM-molecular interface allows transmission  of information under different biological scenarios, where the EM force can be regarded as a powerful tool that controls the energy pathways of proteins.     

\subsection{Capacity Result with Average Energy Constraint}
For the multi-state protein model formulated in Sec.~\ref{Sec5}, we opt  to find the capacity by which a cascade of protein configurations transduce information and carries out interactions upon  THz stimulation. This scenario sparks a resemblance of   enzymes  and  receptors that are activated via protein  phosphorylation. In addition, the selectivity provided by using a THz nanoantenna allows us to control $\alpha_{\mathbf{AF}}$ by governing $p_{\mathbf{AF}}$ applied to each link and therefore bias our network in a specific direction. The  constrained BA  algorithm is deployed, where an average energy constraint is applied to the capacity as formulated in Sec.~\ref{Sec5}-C.  For simulations, we will use the model illustrated in Fig.~\ref{fig:model_2}, constituting of 4 protein states. We  examine different values of $\alpha_{\mathbf{NF}}$ while assuming $\alpha_{\mathbf{NF_1}}=\alpha_{\mathbf{NF_2}}=\alpha_{\mathbf{NF_3}}$. The value of $\beta$ is studied when it is either fixed or varied for the three links. By selecting different values of $\beta$, we can analyze how forward transition rates are impacted as nanoantenna force is being applied to the system. 
\subsubsection{ Fixed $\beta$}
\begin{figure}[h!]
\centering
\includegraphics[width=0.4\textwidth]{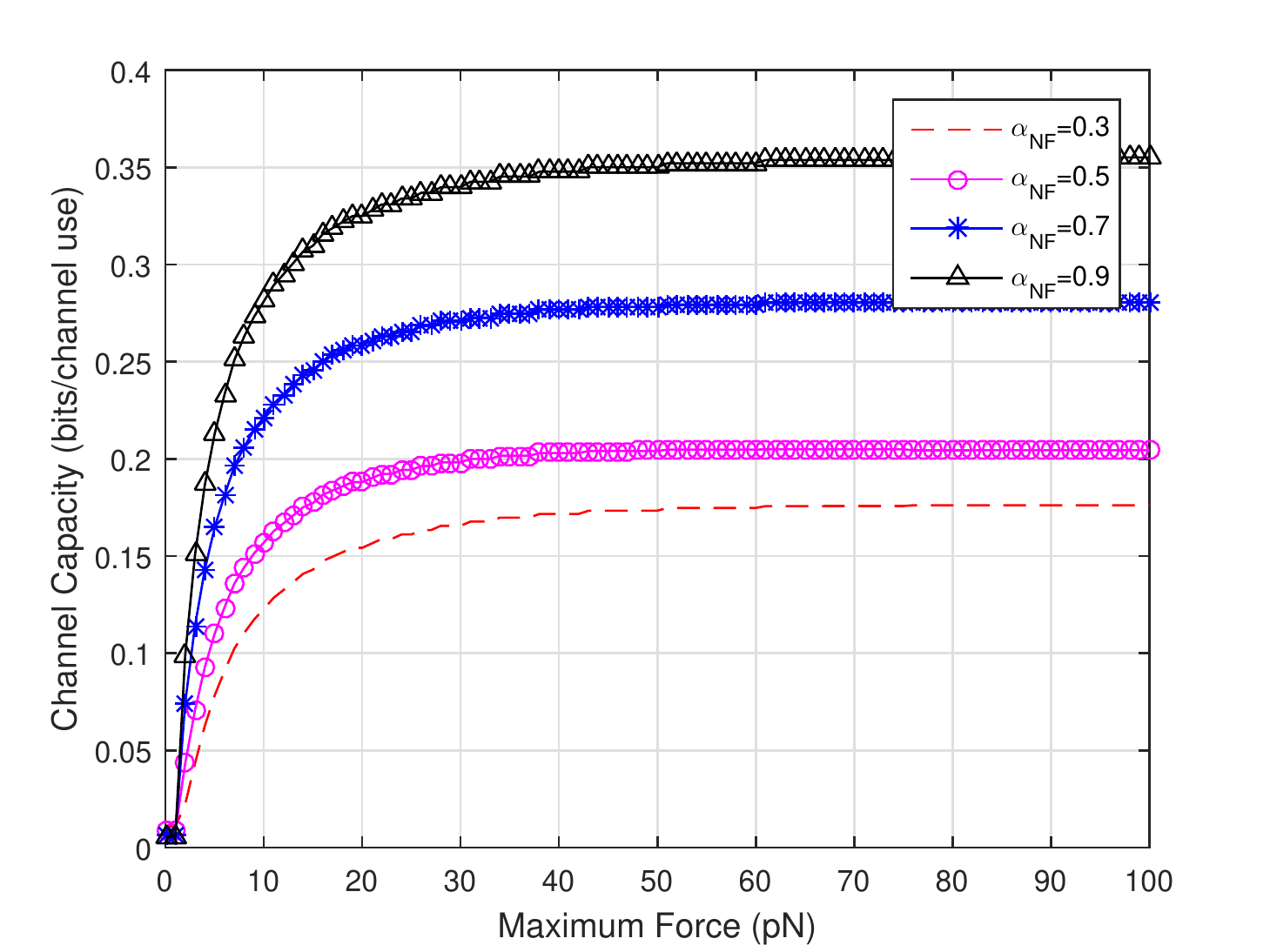}
\footnotesize
\caption{The channel capacity for the multi-state protein model as a function of the nanoantenna applied force. The value of $\beta$ is fixed to $0.9$ for the three links while the value of $\alpha_{\mathbf{NF}}$ varies.} 
\label{fig:model2}
\end{figure}

\begin{figure}[h!]
\centering
\includegraphics[width=0.4\textwidth]{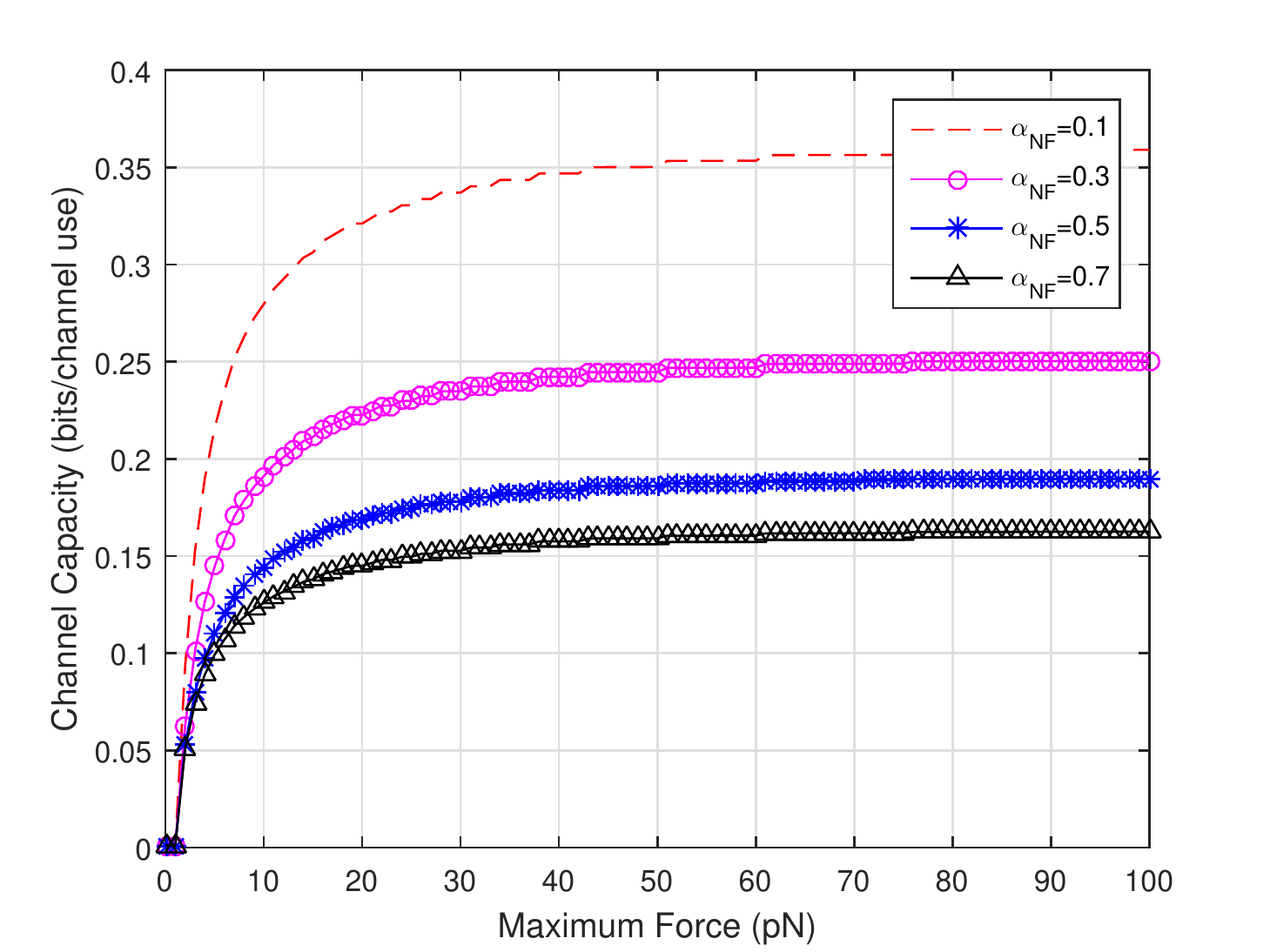}
\footnotesize
\caption{The channel capacity for the multi-state protein model as a function of the nanoantenna applied force. The value of $\beta$ is fixed to $0.1$ for the three links while the value of $\alpha_{\mathbf{NF}}$ varies.} 
\label{fig:model22}
\end{figure}

 Since protein interaction reflects a biological phenomenon, a protein network will favor the condition which achieves equilibrium. As such, at equilibrium, the system will always have the highest capacity as indicated by Figs.~\ref{fig:model2} and~\ref{fig:model22}. The results match the conclusion achieved in Sec.~\ref{Sec7}-B, indicated by~\eqref{eq:concluded_relation}. When the system is out of equilibrium, heat dissipation occurs and work should be done to bring the system back to equilibrium, therefore reducing  the attained capacity. It can be also noticed that the maximum  achieved capacity of Figs.~\ref{fig:model2} and~\ref{fig:model22} is lower compared to Fig.~\ref{fig:cont1}. This is attributed to the energy constraint set by $E^{max}$ in~\eqref{eq:const2}. The chosen $E^{max}$ value corresponds to the typical energy consumed by a motor protein~\cite{howard2001mechanics}.

\subsubsection{Different $\beta$}
\begin{figure}[h!]
\centering
\includegraphics[width=0.4\textwidth]{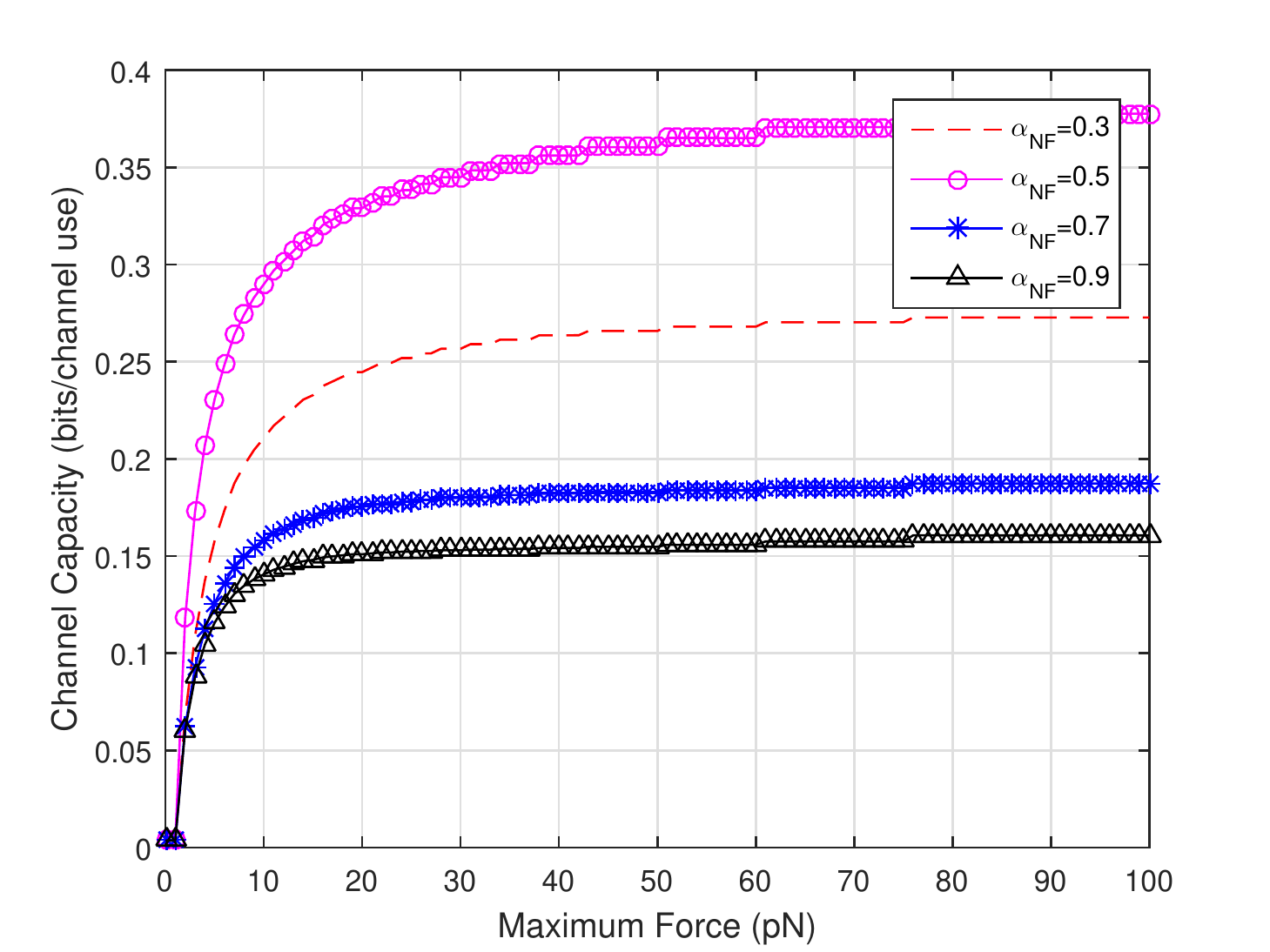}
\footnotesize
\caption{The channel capacity for the multi-state protein model as a function of the nanoantenna applied force. The value of $\beta$ is different for each link where $\beta_1=0.5$, $\beta_2=0.6$, $\beta_3=0.2$.} 
\label{fig:mixedb1}
\end{figure}

\begin{figure}[h!]
\centering
\includegraphics[width=0.4\textwidth]{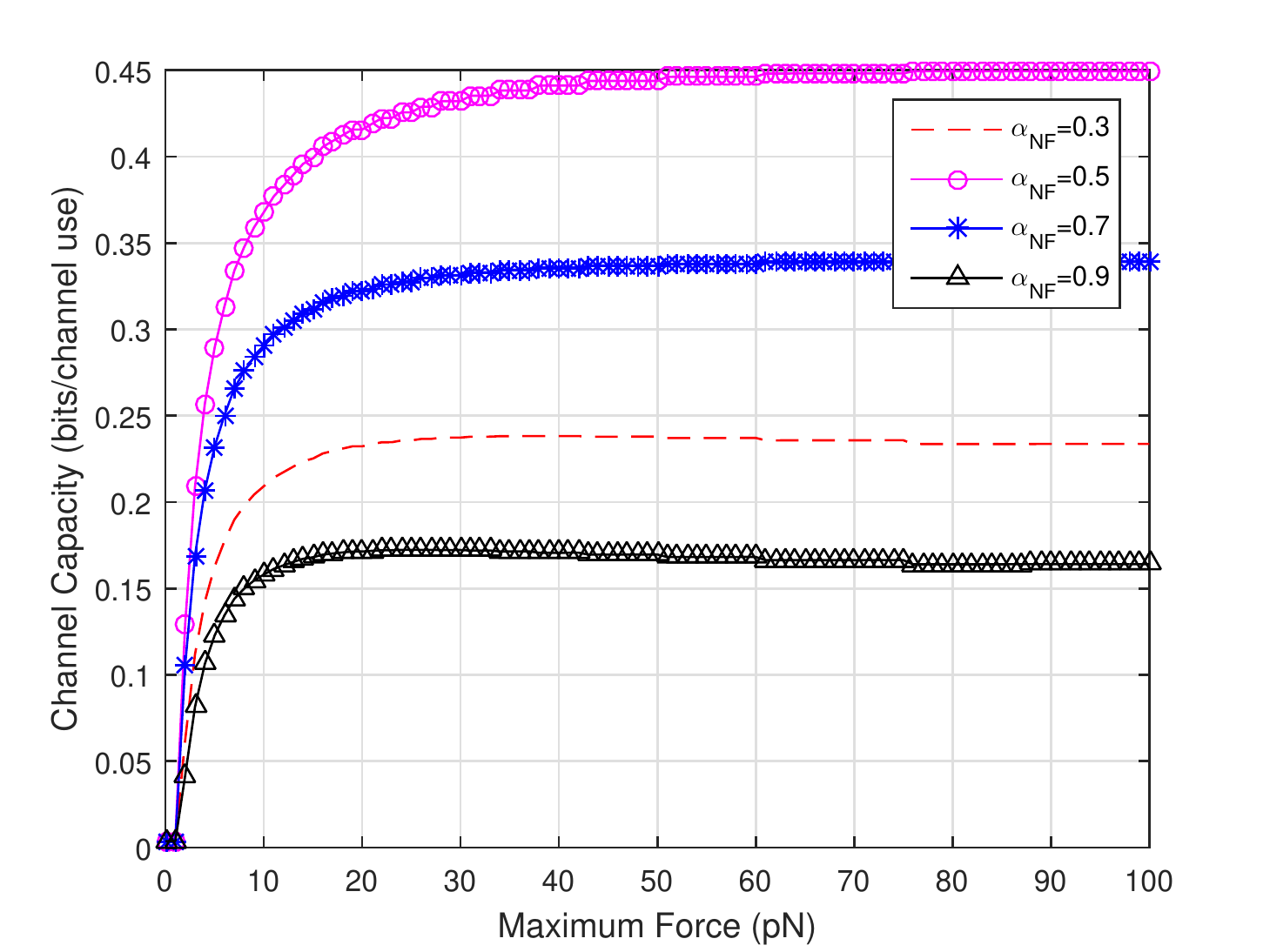}
\footnotesize
\caption{The channel capacity for the multi-state protein model as a function of the nanoantenna applied force. The value of $\beta$ is different for each link where $\beta_1=0.3$, $\beta_2=0.5$, $\beta_3=0.7$.} 
\label{fig:mixedb2}
\end{figure}

 Figs.~\ref{fig:mixedb1} and~\ref{fig:mixedb2} show the channel capacity for the multi-state protein model as a function of the nanoantenna force when the value of $\beta$ is set different for each link. The capacity of the system depends on the combination of $\beta$ and $\alpha_{\mathbf{NF}}$ for the three links as reflected from the mutual information formula. The maximum capacity is achieved when the overall free energy values of the system, composed in our case of the three links, is closest to equilibrium. This relationship is deduced from~\eqref{eq:concluded_relation} and is given as
 \begin{equation}
\Delta E_{ij}^0=k_{b}T \sum_{k=1}^{m} \ln\left[ \frac{\alpha_{\mathbf{NF}_k}}{\beta_k}\right].
\label{eq:concluded_relation2}
\end{equation}
 
 This case resembles a more realistic intra-body scenario because unfolding rates between protein intermediates are not necessarily equal. Our results match the fact that physical systems in equilibrium have a statistical tendency to reach states of maximum entropy or minimum Gibbs free energy~\cite{rietman2016thermodynamic}.

\section{Conclusion and Discussion}
\label{Sec8}
In this paper, we present a communication system which bridges the link between EM nanonetworks and molecular paradigms. The developed stimuli-responsive system constituting of a nanoantenna transmitter and a protein receiver, paves the way towards controlled intra-body interactions at a molecular level. The key idea relies on stimulating the protein vibrational modes to induce a change in their state. Protein conformational changes activate biochemical events that transduce through intra-body pathways.
  
The presented mathematical model  uses the Boltzmann distribution to represent the system states. For the communication channel, a Markov chain finite-state model is used to represent the system inputs and outputs. Both a two-state and a multi-state protein model are developed. In the former model, the focus is on a single folding and unfolding interaction which results in a controlled biological change in the medium followed by a cascade of reactions. Such a model is inspired from mechanosensitive channels that adopt two fundamental conformational channel states separated by an energy barrier. 

In the latter model, we investigate a series of interactions representing a protein undergoing intermediate changes in configuration, where we generalize the presented two-state model. Expressions for the mutual information are derived for both cases, indicating the possible information rates achieved by stimulating proteins by THz nanoantennas. Several capacity constraints are also introduced to make sure the system is compatible with the intra-body medium.

The results attained indicate a feasible communication platform for information transmission between the nanoantenna and the protein. It also expresses a fundamental link between  kinetics and thermodynamics since protein interactions  favor conditions of equilibrium even when an external force is applied to the system, which shows that the results adhere to the second law of thermodynamics. The results agree with the fact that a time-homogeneous Markov chain converges to the Gibbs equilibrium measure, i.e., thermal equilibrium. In essence, the concept of mutual information introduced in this work not only indicates the amount of information the protein signaling pathway carries but can also be further interpreted in terms of molecular disorder, where the highest capacity is obtained when minimum energy is lost. Such a conclusion will result in various medical opportunities where proteins are controlled and directed towards certain favorable interactions. 

As a future direction, we aim to present a mathematical model that captures the interaction between THz  waves and protein dynamics   from a  mechanical perspective. This involves studying the resonance response associated with protein conformational changes  by  modeling the protein as a large set of coupled harmonic oscillators. The mechanical model must be integrated with the current work in order to have a complete system that relates the triggered natural frequencies of proteins to the probability of folding. In addition, the authors would  like to further study the relationship between THz waves and misfolded proteins associated with neurodegenerative diseases. This involves understanding how THz waves may alter the pathological  mechanisms  and  how  this  knowledge  can be reflected to develop   disease-modifying therapeutic strategies.     

\bibliographystyle{IEEEtran}
\bibliography{IEEEabrv,references}

\begin{IEEEbiography}
[{\includegraphics [width=1.1 in,height=1.3 in,keepaspectratio]{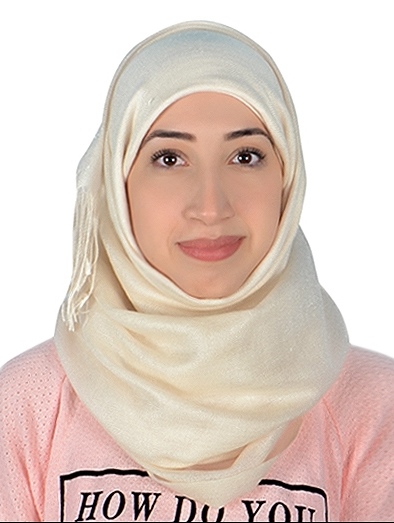}}]{Hadeel Elayan}(S'12)  is currently a  PhD Candidate in the Electrical and Computer Engineering department at the University of Toronto, Canada. Her  research interests include Nanonetworks, Terahertz  Intra-body Communication as well as Molecular Communication.   Hadeel completed a research internship at the Ultra-broadband Nanonetworking Lab, University at Buffalo, USA during summer 2016. She worked as a Research Associate in the Healthcare Engineering Innovation Center, Khalifa University until August 2018. Hadeel received several awards for her research and academic excellence including  the   2016 IEEE Pre-doctoral Research Grant Award,  the 2017 Photonics School Internship Award from KAUST  and the 2019 Ontario Graduate Scholarship.\end{IEEEbiography}
 
 \begin{IEEEbiography}
[{\includegraphics [width=1.1 in, height=1.3in]{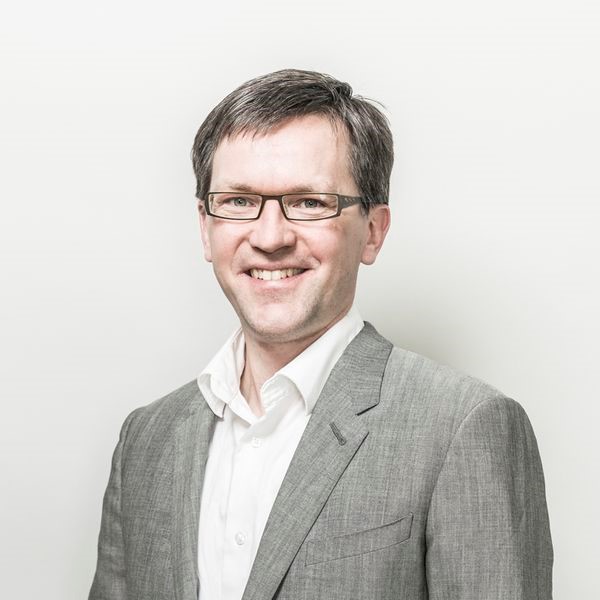}}]{Andrew Eckford}is an Associate Professor in the Department of Electrical Engineering and Computer Science at York University, Toronto, Ontario. His research interests include the application of information theory to biology, and the design of communication systems using molecular and biological techniques. His research has been covered in media including The Economist, The Wall Street Journal, and IEEE Spectrum. His research received the 2015 IET Communications Innovation Award, and was a finalist for the 2014 Bell Labs Prize. He is also a co-author of the textbook Molecular Communication, published by Cambridge University Press.\end{IEEEbiography}

\vspace{-250 pt}

\begin{IEEEbiography}
[{\includegraphics [width=1.1 in,height=1.3 in]{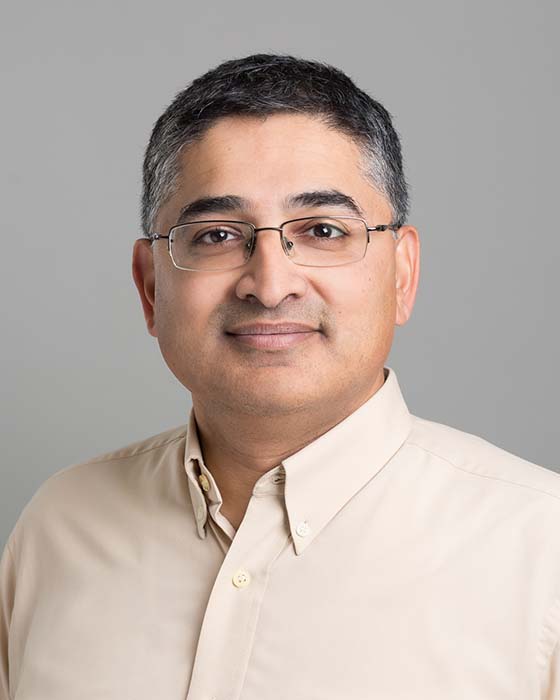}}]{Raviraj S. Adve} was born in Bombay, India. He received the B.Tech. degree in
Electrical Engineering from IIT Bombay in 1990 and the Ph.D. from Syracuse
University in 1996, where his thesis won the Syracuse University Outstanding
Dissertation Award. From 1997 to 2000,  he was with Research Associates for
Defense Conversion Inc., on contract with the Air Force Research Laboratory,
Rome, NY, USA. He joined the Faculty of the University of Toronto in 2000,
where he is currently a Professor. His research interests include molecular
communications, analysis and design techniques for cooperative and
heterogeneous networks, energy harvesting networks, and in signal processing
techniques for radar and sonar systems. He received the 2009 Fred Nathanson
Young Radar Engineer of the Year Award. He is a fellow of the IEEE.\end{IEEEbiography}
\end{document}